\documentclass[paper=letter,american,fontsize=11pt,DIV=10]{scrartcl}
\usepackage{times}
\usepackage[T1]{fontenc}
\usepackage[utf8]{inputenc}
\usepackage{hyphenat}
\usepackage{babel}
\usepackage{amsmath}
\usepackage{amssymb}
\usepackage{amsthm}
\usepackage{xfrac} 
\usepackage{verbatim}
\usepackage[normalem]{ulem}

\usepackage{graphicx}	
\usepackage{xcolor}

\usepackage[authoryear,round]{natbib}
\usepackage{latexsym}
\usepackage{authblk} 
\usepackage[unicode=true,bookmarks=true,bookmarksnumbered=false,bookmarksopen=false, breaklinks=false,pdfborder={0 0 0}, backref=false,colorlinks=false]{hyperref}   

\newtheorem*{theorem*}{Theorem}

\begin{document}
\pagenumbering{roman}

\title{Is the Statistical Interpretation of \\
Quantum Mechanics 
$\psi$-Ontic or $\psi$-Epistemic?}
\author{Mario Hubert}
\affil{The American University in Cairo}


\date{November 13, 2022 \\
\vspace{1.5 em}
{\Large Forthcoming in  \emph{Foundations of Physics
}}}
\maketitle
\begin{abstract}
The ontological models framework distinguishes $\psi$-ontic from $\psi$-epistemic wave-functions. It is, in general, quite straightforward to categorize the  wave-function of a certain quantum theory. Nevertheless, there has been a debate about the ontological status of the wave-function in the statistical interpretation of quantum mechanics: is it $\psi$-epistemic and incomplete or $\psi$-ontic and complete? I will argue that the wave-function in this interpretation is best regarded as  $\psi$-ontic \emph{and incomplete}.    
\end{abstract}

\newpage

\tableofcontents

\newpage

\pagenumbering{arabic}

\section{Introduction}

In their recent article, \citet{Oldofredi:2020wv} claim that \citet{Harrigan:2010aa} define the ontic state $\lambda$ in too restrictive a way so that it does not properly capture the statistical interpretation of quantum mechanics. Furthermore, they argue that \citet{Harrigan:2010aa} put the wave-function in the statistical interpretation in the wrong category: instead of being $\psi$-epistemic (and incomplete) the wave-function is $\psi$-ontic (and complete). They support their argument by historic evidence from Albert Einstein and Leslie Ballentine, who are the most prominent proponents of the statistical interpretation. I want to challenge all three claims and argue for the following: (i) the ontic state in the sense of \citet{Harrigan:2010aa} is general enough to capture a statistical interpretation of the wave-function, (ii) the wave-function can be interpreted as $\psi$-epistemic as perceived by Einstein and Ballentine (although it does not merely represent an observer's knowledge), and (iii) the statistical interpretation is more compatible with an ontic state as described by \citet{Harrigan:2010aa} than as re-defined by \citet{Oldofredi:2020wv}.

I will argue that both interpretations face different kinds of problems: the interpretation of \citet{Harrigan:2010aa} runs into the PBR-theorem, and the interpretation of \citet{Oldofredi:2020wv} is in fact incomplete itself and conceptually unclear. Ultimately, the status of the wave-function in the statistical interpretation depends on the particular completion of this theory. Yet, I will argue that the wave-function in the statistical interpretation is most plausibly $\psi$-ontic and incomplete. 




 \section{The Ontological Models Framework}

Let me first explain where the notion of $\psi$-epistemic comes from and what it means. \citet{Harrigan:2010aa} introduce it within a certain formal framework for quantum mechanics, which they call the \emph{ontological models framework}. The main question they ask is, ``Does the quantum state represent reality or only our knowledge of reality?'' One option is to sidestep this question and take an operational approach to quantum mechanics: all one is interested in is predicting the behavior of quantum systems without recourse to unobservable objects and processes. The ideal case would be an \emph{operational quantum theory} that describes all predictions of the theory in terms of preparation and measurement procedures.  In doing so, it needs to say what kinds of measurements $M$ and preparation procedures $P$ yield an outcome $k$. An operational quantum theory specifies a probability $\mathbb{P}$ for $k$, given $M$ and $P$, that is, $\mathbb{P}(k\vert M,P)$. 

Often a physical theory tells us more about the world than how certain manipulations on a physical system lead to certain empirical results: it may tell us what the measured system is made of, what a measurement device is made of, and how measurement and preparation devices interact with the constituents of the measured system \citep{Hubert:2020ab}.  In particular, the physical theory should (ideally) specify a complete description of the system's properties. This complete description is often denoted $\lambda$. A theory that gives such an ontological story to operational quantum theory is an \emph{ontological model of operational quantum theory}.

Again, an ontological model of an operational theory of quantum mechanics provides more information about the physical system beyond an operational level; it describes the complete physical state $\lambda$ of the system. Although measurement and preparation devices can be described with their own ontic state $\lambda$ in an ontological theory,  they are not reduced to these $\lambda$'s in the ontological models framework. The main problem that the ontological models framework aims to tackle is how much an observer can know about the ontic state of the physical system if she prepares and measures the system in a certain way. For example, a preparation procedure may not uniquely fix the ontic state but rather a probability distribution $\mathbb{P}(\lambda \vert P)$ for ontic states given a preparation procedure. If the observer prepares the system in a state $\lambda$, this ontic state determines then the result $k$ once measured, that is, the ontological models framework provides this probability distribution $\mathbb{P}(k \vert \lambda, M)$. We can now explain the predictions of the operational theory $\mathbb{P}(k\vert M,P)$ with the machinery of the ontological theory: 
$$\mathbb{P}(k\vert M,P)=\int  \mathrm{d}\lambda \, \mathbb{P}(k \vert \lambda, M) \mathbb{P}(\lambda \vert P).$$
We integrate over all the possible ontic states, where $\mathbb{P}(\lambda \vert P)\neq0$ singles out the ontic states that are compatible with the state procedure $P$.

It is often implied that the ontological model framework is a strong constraint on a physical theory \citep[for instance,][]{Gao:2017aa}. In my opinion, it is rather a formalization of how physics has been ever done until recently with the development of quantum physics by identifying the complete physical state of a system and investigating the time evolution of this state. One may still agree with me here and still claim that the ontological models framework is too restrictive for quantum physics (I thank an anonymous reviewer for this remark). Nevertheless, the ontological models framework reaches a large set of quantum theories that does not face the physical, conceptual, and ontological problems as theories that do not fit into this model \citep[see, for instance,][]{Chen:2021vv,Norsen:2016aa}.  

\citet[][]{Oldofredi:2020wv} seem to mischaracterize the relation between an operational theory and its corresponding ontological model:
 
\begin{quote}
It is worth noting that the authors define ontological models employing an operational setting, i.e.\ the primitive notions of such models consist exclusively in preparations procedures of physical systems in certain states and measurements performed on them. A complete specification of the properties of a given physical system is provided by $\lambda$, the ontic state of the system under scrutiny. \citep[][pp.\ 1318--1319]{Oldofredi:2020wv}
\end{quote}

\begin{quote}
As we have seen, ontological models consider operational procedures to prepare the state of a quantum system in a certain manner as primitive notions. Such procedures are associated with some observable properties, whose values will be then revealed by the performance of a set of measurements on the physical system under scrutiny. \citep[][pp.\ 1321]{Oldofredi:2020wv}
\end{quote}
 
 It is not the ontological model that treats the operational procedures as primitive notions. Rather, the ontological model provides the tools for \emph{explaining} these procedures and \emph{reduce} them to the behavior of the ontic state $\lambda$. More precisely, ''In an ontological model of an operational theory, the primitives of description are the properties of microscopic systems,'' \citep[][p.\ 128]{Harrigan:2010aa} and these properties encoded in $\lambda$ determine the probabilities and outcomes in preparation and measurement procedures. Of course, the ontological model itself does not tell us exactly how the operational procedures look like, but it demands, in contrast to an operational theory, that the probabilities derive from the interaction with the ontic state of the system. It is only on the level of the operational quantum theory, where the operational procedures are taken as primitive. Therefore, the relation between the operational theory and the corresponding ontological model is indeed similar to the relation between thermodynamics and statistical mechanics.
 

In quantum mechanics, the wave-function has a double role: it determines the probabilities of measurement outcomes via the Born rule, and it also (at least partially) describes the ontic state of the physical system. This is also reflected in the ontological models framework. The probabilities of the operational theory are determined by the wave-function $\mathbb{P}(k\vert M,P)=\mathbb{P}_{\psi}(k\vert M,P)$. The ontological models framework zooms in on the relation between the ontic state $\lambda$ and the corresponding wave-function describing this state. If the system is prepared to have a certain wave-function $\psi$, the system may be in one of many possible ontic states compatible with this $\psi$. The corresponding probability distribution would be $\mathbb{P}(\lambda \vert P=\psi)$.


Having introduced the ontic state $\lambda$, one can then try to answer the original question ``Does the quantum state represent reality or our knowledge of reality?''. In other words, does the wave-function represent objective properties of the ontic state (the complete physical description of the system), or does it rather represent an agent's knowledge about properties of this state? 
 
\citet{Harrigan:2010aa} introduce an ingenious and rather general formal distinction to make this question more precise. 
If we prepare two systems with different wave-functions, a natural question arises of how the ontic states corresponding to these wave-functions are related. The ontological models framework distinguishes between two cases:
\begin{enumerate}
\item[(i)]
If the probability distributions $\mathbb{P}(\lambda \vert P=\psi_A)$ and $\mathbb{P}(\lambda \vert P=\psi_B)$ of two different wave-functions $\psi_A$ and $\psi_B$ do not overlap, as depicted in Fig.\ \ref{fig:psi-ontic-epistemic}a), the wave-function is called \emph{$\psi$-ontic}.
 \begin{figure}[ht]
 \centering
 \includegraphics[width=15.5cm]{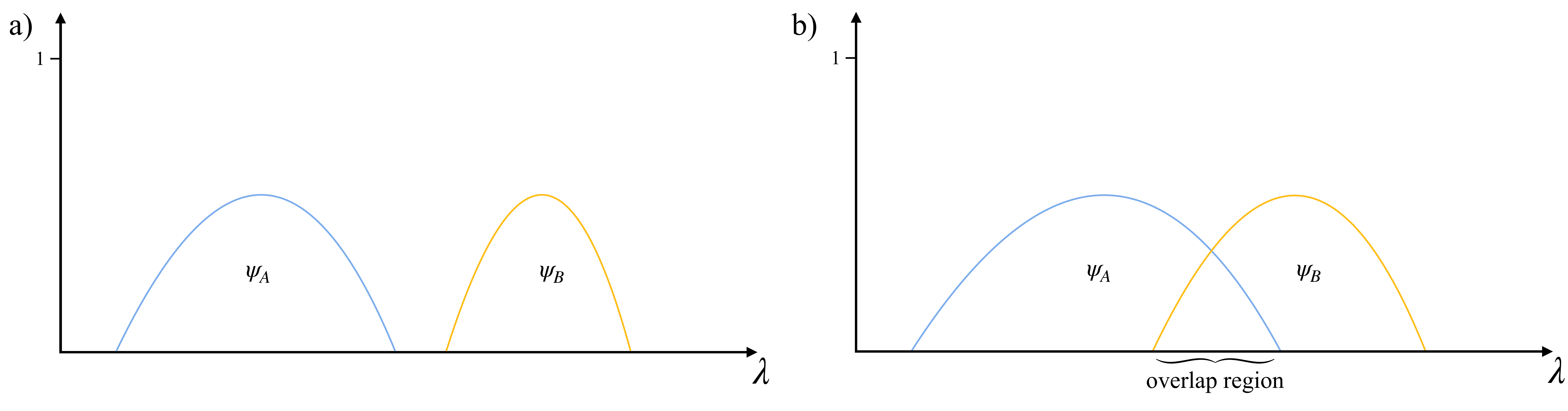}
\caption{a) The definition of a $\psi$-ontic wave-function: The probability distributions over the ontic states associated with the wave-functions $\psi_A$ and $\psi_B$ are disjoint. b) The definition of a $\psi$-epistemic wave-function: The probability distributions over the ontic states associated with the wave-function $\psi_A$ and $\psi_B$ share a region of overlap.}
\label{fig:psi-ontic-epistemic}
\end{figure}
\item[(ii)]
If the probability distributions $\mathbb{P}(\lambda \vert P=\psi_A)$ and $\mathbb{P}(\lambda \vert P=\psi_B)$ of two different wave-functions $\psi_A$ and $\psi_B$ do overlap, as depicted in Fig.\ \ref{fig:psi-ontic-epistemic}b), the wave-function is called \emph{$\psi$-epistemic}.
\end{enumerate}

For every wave-function, there is a set of ontic states $\lambda$ compatible with this wave-function, that is, there is a one-to-many relation from the wave-function to the underlying ontic states. If we pick an ontic state, then it depends on the wave-function whether the relation from the ontic state to the wave-function is one-to-one or one-to-many. More precisely, if the wave-function is $\psi$-ontic, fixing an ontic state $\lambda$, there is only one wave-function associated with it. If the wave-function is $\psi$-epistemic, there are ontic states which are associated with at least two wave-functions---namely, the ones in the overlap region.

The definition of $\psi$-ontic and $\psi$-epistemic was up to now only about whether or not a wave-function can be uniquely associated with an ontic state. On this level, $\psi$-ontic and $\psi$-epistemic are merely formal notions. Nevertheless, these definitions have been introduced to clarify and make more precise in what way the wave-function is ontic or epistemic; that is, in what way the wave-function represents objective features or knowledge about the world. But we will shortly encounter certain caveats regarding the ontological and epistemic status of $\psi$-ontic and $\psi$-epistemic wave-functions. $\psi$-ontic and $\psi$-epistemic carve up the ontic and epistemic landscape of the wave-function a bit differently than the notions may suggest. 

\subsection{How Ontological Are $\psi$-Ontic Wave-Functions?}

Let us first discuss in which way a $\psi$-ontic wave-function is ontological. If an ontic state is uniquely associated with a wave-function, then and only then can we interpret the wave-function as representing only certain objective properties of $\lambda$ (what these properties exactly are is left open). 

\begin{figure}[ht]
  \centering
 \includegraphics[width=14cm]{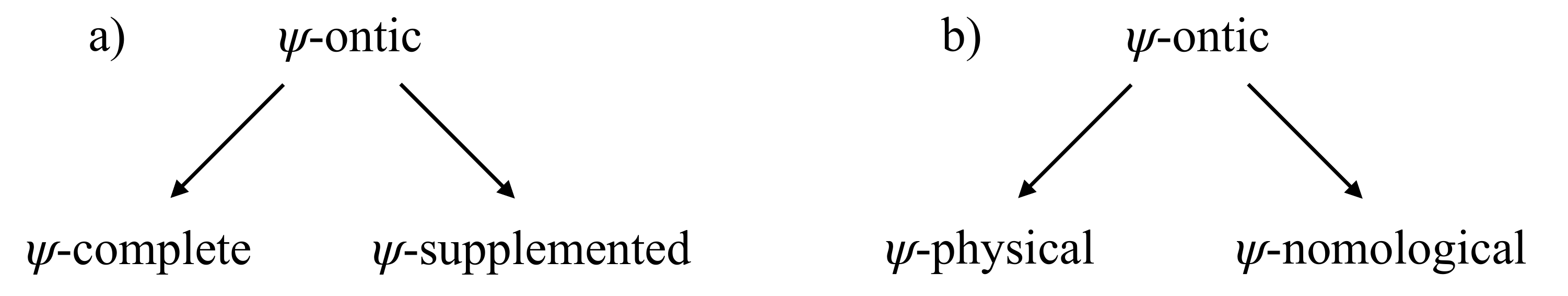}
\caption{a) $\psi$-ontic wave-function can be either $\psi$-complete or $\psi$-supplemented. A $\psi$-complete wave-function describes completely the ontic state $\lambda$. A $\psi$-supplemented wave-function only gives a partial description. b) A $\psi$-ontic wave-function can be either a physical object ($\psi$-physical) or an abstract object representing properties of the system ($\psi$-nomological).}
\label{fig:psi-ontic-versions}
\end{figure}

\begin{figure}[ht]
  \centering
 \includegraphics[width=10cm]{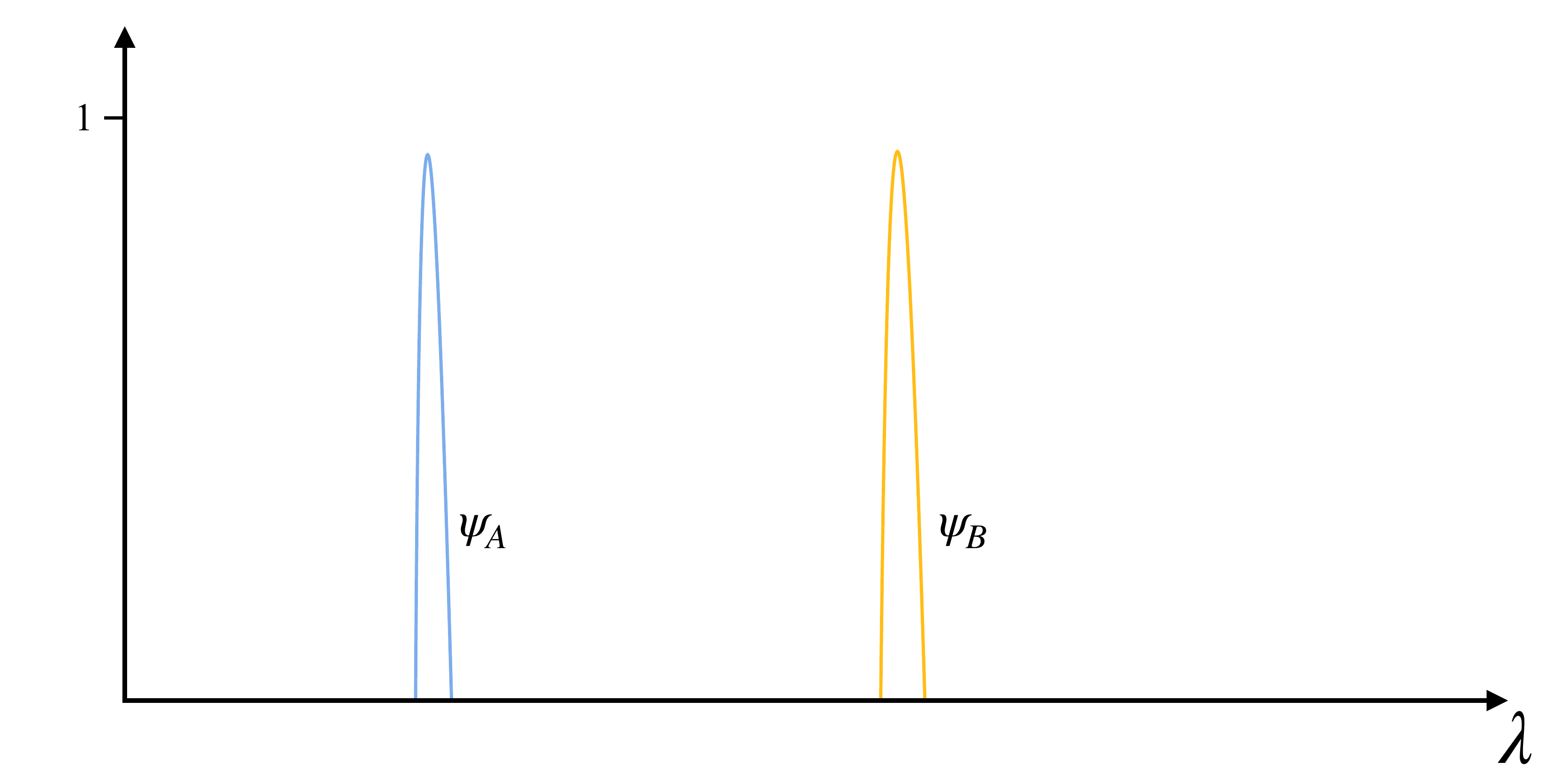}
\caption{$\psi$-complete wave-function. The wave-functions $\psi_A$ and $\psi_B$ are narrowly peaked around an ontic state $\lambda$.}
\label{fig:psi-complete-peaked}
\end{figure}

We can introduce further definitions that show how a $\psi$-ontic wave-function can objectively represent these properties (see Fig.\ \ref{fig:psi-ontic-versions}a). First, the wave-function alone can completely describe the state of the system. In this case, we call the wave-function $\psi$-complete. The probability distributions of $\psi$-complete wave-functions are sharply peaked around $\lambda$, as shown in Fig.\ \ref{fig:psi-complete-peaked}.
If the wave-function does not completely describe $\lambda$, then it is called $\psi$-supplemented. A $\psi$-supplemented wave-function would need additional variables to provide such a complete description (what these variables are is also left open in the ontological models framework).

We can also describe $\psi$-ontic wave-functions in a different way, as done in Fig.\ \ref{fig:psi-ontic-versions}b).
If the wave-function represents properties of $\lambda$ because it is itself a physical object (or rather the object it represents is a physical object), then we call it $\psi$-physical. If the wave-function represents an abstract entity that still determines the behavior of $\lambda$, we call it $\psi$-nomological.\footnote{An anonymous reviewer challenged my distinction, in the following way (I thank the reviewer for this argument). First, \citet{Harrigan:2010aa} had in mind a $\psi$-physical wave-function when defining a $\psi$-ontic wave-function; therefore, $\psi$-nomological wave-functions cannot be $\psi$-ontic. I could not retrieve this presupposition in Harrigan and Spekkens' paper. Harrigan and Spekkens are rather indifferent to the metaphysical status of the wave-function. Instead, they propose a \emph{formal} relationship between the ontic state and the wave-function elucidating how the properties of a physical system are represented by the wave-function. \citet[][p.\ 8]{Maudlin:2022aa} also regards nomological wave-functions as examples of $psi$-ontic wave-functions and mentions even more subcategories of $\psi$-ontic wave-functions than I do. Second, if the the wave-function is $\psi$-nomological it cannot represent the physical properties of the ontic state $\lambda$ qua abstract entity. I suppose that underlying this argument is too strict a reading of the word ''physical'' in the definition of the reality criterion. A $\psi$-nomological wave-function can still represent parts of the complete physical properties of the quantum system, even if it is an abstract nomological entity. The wave-function does not need to be a physical object, like a field, to be included in the physical properties of the quantum system.}

\begin{figure}[ht]
  \centering
 \includegraphics[width=15cm]{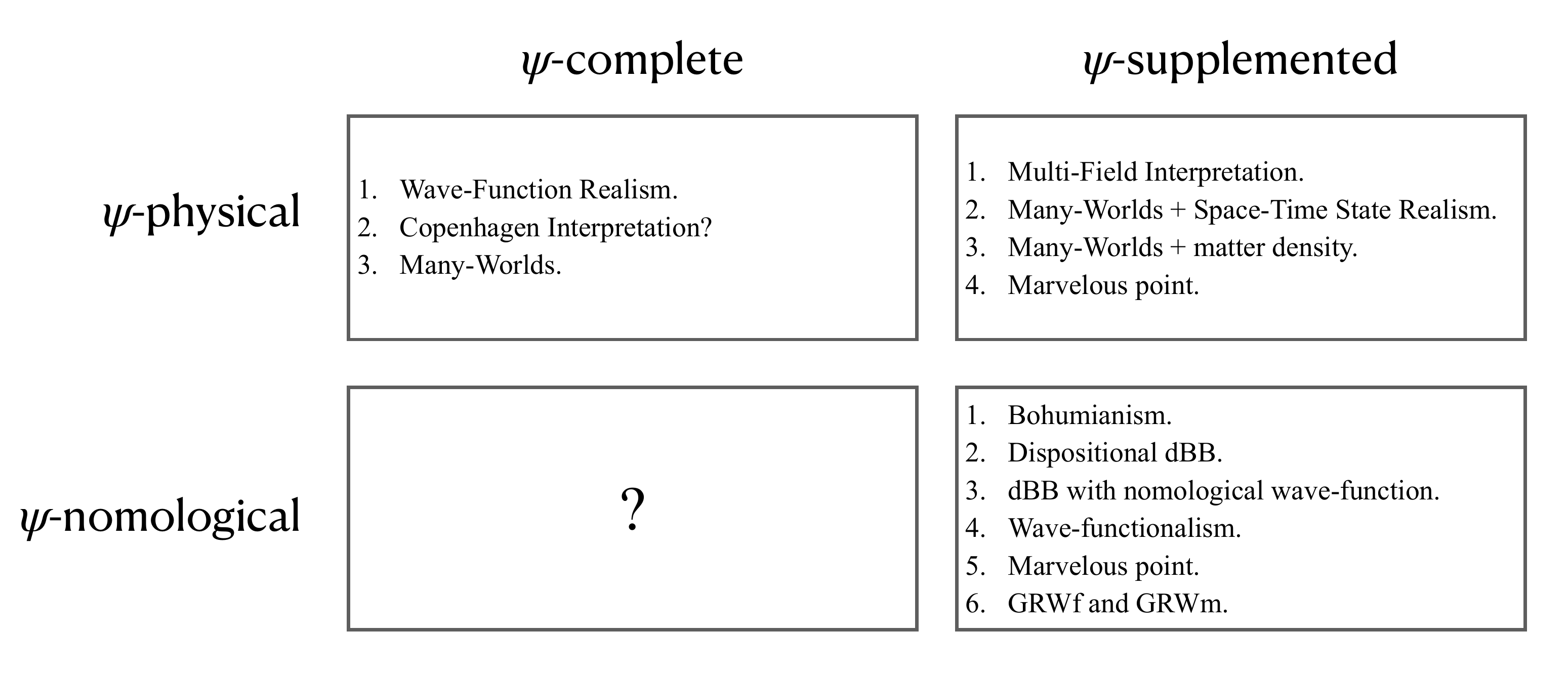}
\caption{The many examples of a $\psi$-ontic wave-function.}
\label{fig:psi-matrix}
\end{figure}

We can now build a matrix for the different combinations of $\psi$-ontic wave-functions (see Fig.\ \ref{fig:psi-matrix}), as the ontological models framework merely provides a general categorization of the wave-function that needs to be filled in by specific interpretations of quantum mechanics.  
Example of theories or interpretations that construe the wave-function as $\psi$-complete and $\psi$-physical are wave-function realism \citep{Ney:2021vp}, probably the Copenhagen interpretation, and different versions of the many-worlds interpretation \citep{Carroll:2019wp}. Examples of $\psi$-supplemented and $\psi$-physical are the multi-field interpretation within the de Broglie--Bohm theory \citep{Hubert:2018aa,Romano:2021vz}, space-time state realism in the many-worlds theory \citep{Wallace:2003aa}, the many-worlds theory with matter density \citep{Allori:2011aa}, and the original version of Albert's marvelous point interpretation \citep{Albert:1996aa}. Examples of theories or interpretations with a $\psi$-supplemented and $\psi$-nomological wave-function would be  Humean interpretations of the de Broglie--Bohm theory \citep[also called Bohumianism,][]{Esfeld:2012mz,Esfeld:2014ac,Miller:2013aa,Callender:2015aa, Bhogal:2015aa,Dewar:2020ti}, the dispositional interpretation of the wave-function in the de Broglie--Bohm theory \citep{Esfeld:2012mz,Esfeld:2014ac,Suarez:2015aa}, the wave-function as a nomological entity in the de Broglie--Bohm theory \citep{Goldstein:2013aa}, wave-functionalism \citep{Allori:2021tn}, a Humean version of Albert's marvelous point interpretation \citep{Loewer:1996aa}, and the GRW theory with a flash or matter ontology \citep{Dorato:2010aa,Egg:2015ti,Lorenzetti:2021vf}.\footnote{The philosophical literature on the GRW theories seems to imply that the universal wave-function is $\psi$-nomological. It still needs to be worked out what a $\psi$-physical wave-function amounts to for collapse theories.} A $\psi$-nomological interpretation with a $\psi$-complete wave-function has not been proposed, since it is hard to grasp how the wave-function can completely represent the ontic state with it being nomological entity. 

\subsection{How Epistemic Are $\psi$-Epistemic Wave-Functions?}

\begin{figure}[ht]
  \centering
 \includegraphics[width=9cm]{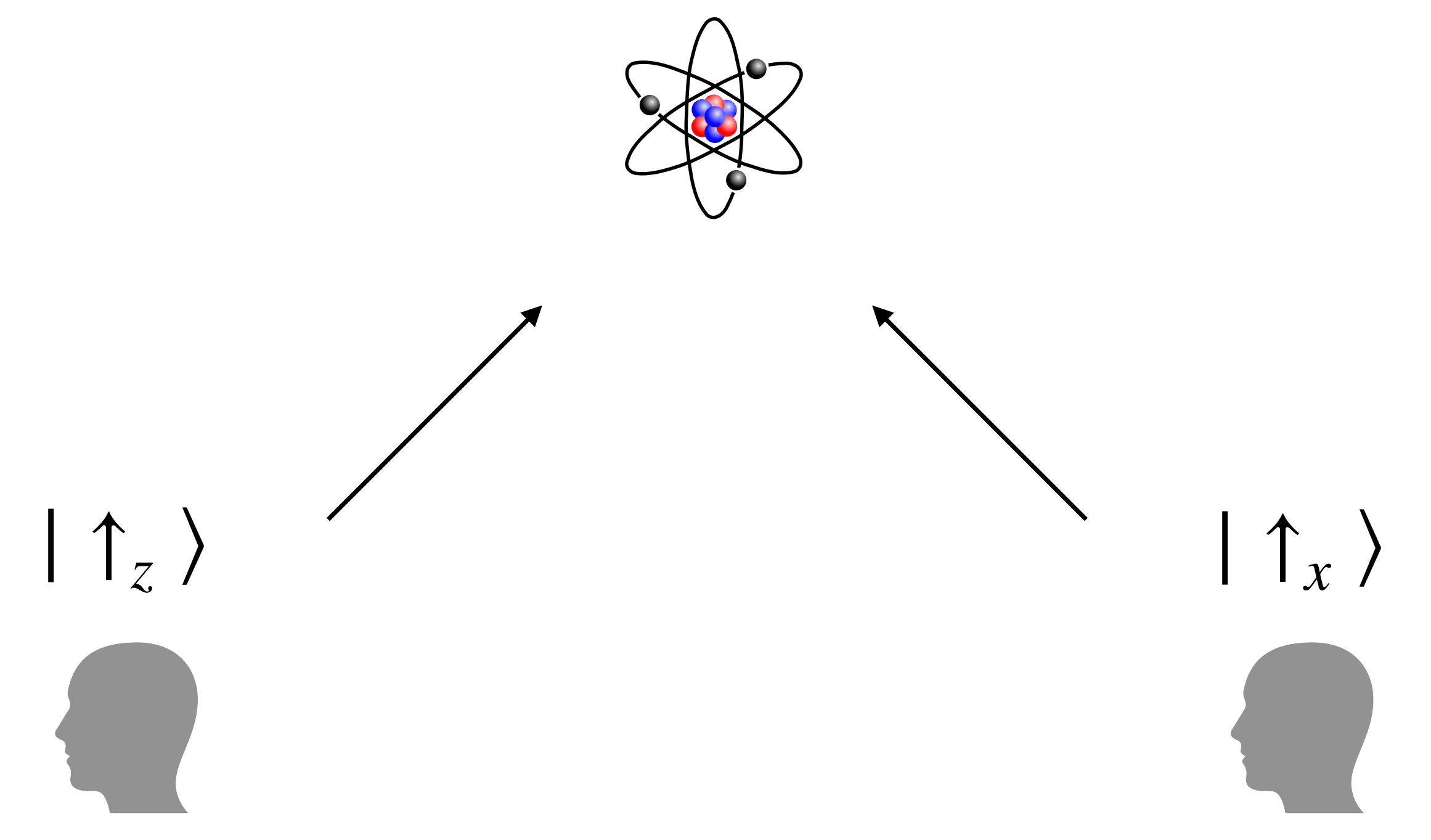}
\caption{One possibility of a $\psi$-epistemic wave-function: Two agents assign two different wave-functions to the same system.}
\label{fig:psi-credal}
\end{figure}

Now let us turn to $\psi$-epistemic wave-functions. These are the ones where some ontic states $\lambda$ are associated with more than one wave-function. How can that happen? One obvious way is when two agents have different knowledge about the same system and disagree on the the wave-function, as in Fig.\ \ref{fig:psi-credal}. It would be a mistake to say that the wave-function in this case is purely epistemic; rather, agents would learn something about the system if they assign to the system one of the (correct) wave-functions that are associated with the ontic state.\footnote{I thank 
Travis Norsen 
for this insight.} This kind of $\psi$-epistemic wave-function emphasizes the relational character between the objective properties of $\lambda$ and the agents' knowledge about $\lambda$ represented in the wave-function. We could call such a wave-function $\psi$-credal, as there is another way to interpret $\psi$-epistemic wave-functions \citep[see][Ch.\ 3]{Maudlin:2019aa}.

\citet[][]{Oldofredi:2020wv} assume that \emph{all} $\psi$-epistemic wave-functions are $\psi$-credal:
\begin{quote}
If a model is $\psi$-epistemic, then it cannot in any case be $\psi$-ontic, since it does not describe any underlying physical reality, but only the agents’ knowledge of it. \citep[][pp.\ 1320]{Oldofredi:2020wv}\footnote{They say the same also on p.\ 1318 and 1328.}
\end{quote}
Even if a $\psi$-epistemic wave-function were to represent an agent's knowledge, the wave-function would be referring to the ontic state. So if I assign a correct $\psi$-epistemic wave-function to a quantum system, I would indeed know some aspects of the underlying physical reality, even if another agent may assign a different wave-function to the same system.

\begin{figure}[ht]
  \centering
 \includegraphics[width=11cm]{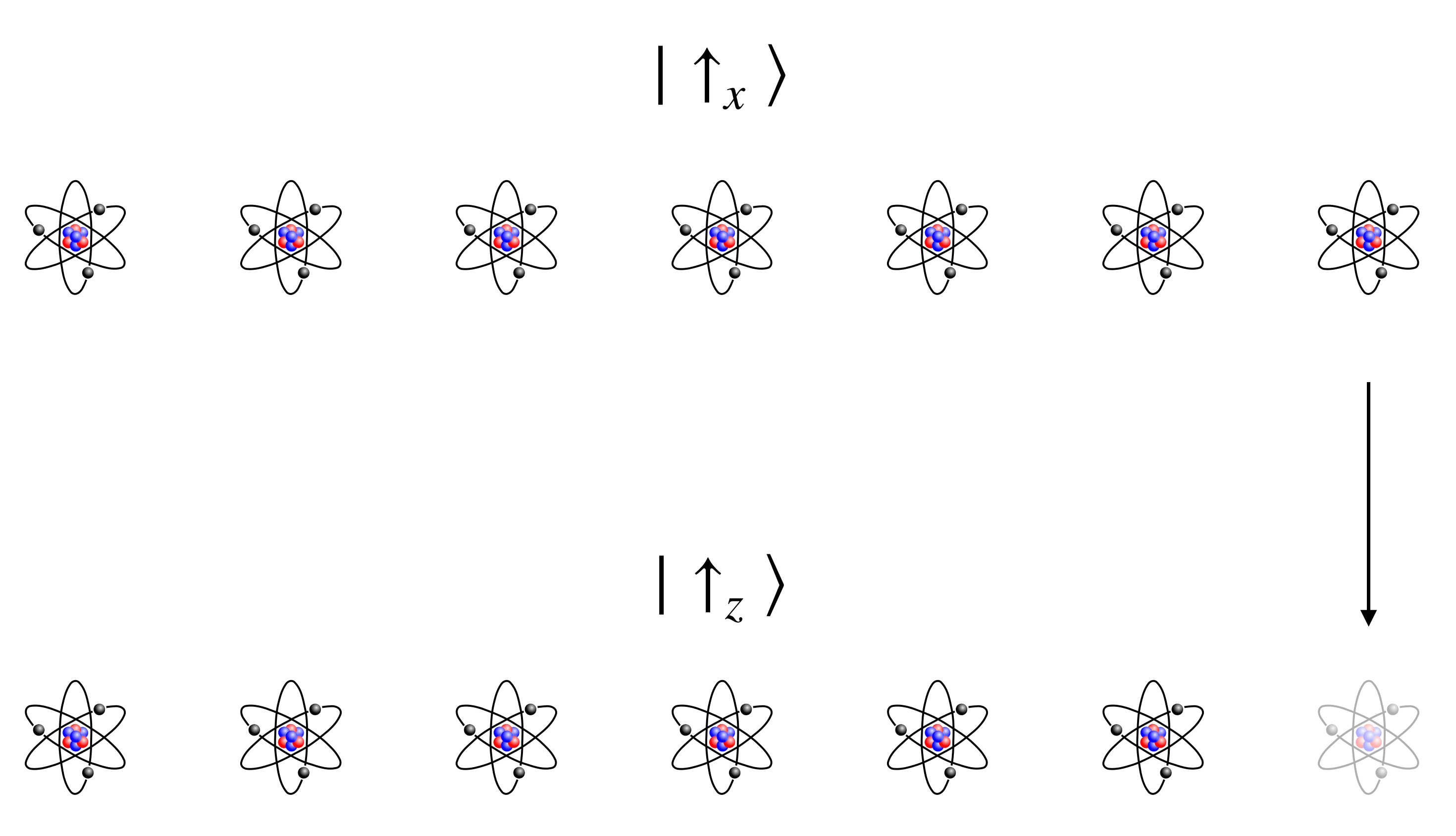}
\caption{The definition of a $\psi$-statistical wave-function: The wave-function describes ensembles of equally prepared quantum systems. In the picture, the upper beam of quantum systems is prepared in a spin x-up state; the lower beam in a spin z-up beam. By definition of $\psi$-epistemic, a certain fraction of quantum systems in each beam can be described by two wave-functions. Therefore, say, the last system in the upper beam can be also correctly associated with the spin z-up ensemble.}
\label{fig:psi-statistical}
\end{figure}

It is indeed possible to interpret a $\psi$-epistemic wave-function in a completely non-epistemic way.\footnote{It is also possible to have an epistemic interpretation of the wave-function without it being $\psi$-epistemic. Therefore, the distinction between $\psi$-ontic and $\psi$-epistemic wave-functions does not cover all the ways one can interpret the wave-function. Pace \citet[][pp.\ 1320]{Oldofredi:2020wv}, the wave-function in QBism would be such an example, as this theory denies the existence of $\lambda$, which is necessary for a wave-function to be $\psi$-epistemic. What the ontology of QBism actually is is still debated \citep[see, for instance,][]{Boge:2021vu}. On the other hand, \citet[][p.\ 1341]{Oldofredi:2020wv} classify relational quantum mechanics to be $\psi$-epistemic  because: ``For RQM, by
contrast, the quantum state is merely a useful tool for calculation and prediction, and
because of this it is $\psi$-epistemic.''}  Imagine we prepare two beams of particles, one with wave-function $\psi_A$ and the other with wave-function $\psi_B$, and we assume that the wave-function describes ensembles. Then whether a particle has wave-function $\psi_A$ is primarily a matter of whether it is part of the ensemble being prepared as having  $\psi_A$ as its wave-function. If the wave-function is $\psi$-epistemic there are some of the particles in the $\psi_A$ ensemble that may be also correctly described by another wave-function, say $\psi_B$. This means that these particles can be correctly associated with  either beam, the $\psi_A$ beam or the $\psi_B$ beam. In this case, the wave-function, although $\psi$-epistemic, does not represent an agent's knowledge but rather whether a particle is part of a certain ensemble. \citet[][Ch.\ 3]{Maudlin:2019aa} calls such a wave-function \emph{$\psi$-statistical} \ (see Fig.\ \ref{fig:psi-statistical}).

This counterexample challenges two arguments by \citet[][]{Oldofredi:2020wv}: (i) that a $\psi$-epistemic wave-function is necessarily epistemic, and (ii) that the wave-function in the statistical interpretation has to be $\psi$-ontic. Having replied to (i), I will discuss (ii) in section \ref{sec:stat-int-ontic}, but before doing so, I shall introduce the PBR-theorem, which will be important to answer more completely whether the statistical interpretation of quantum mechanics can be $\psi$-epistemic. This will in particular affect how \citet{Harrigan:2010aa} understand the statistical interpretation. 

\section{The PBR-Theorem}
\label{sec:pbr-theorem}

The PBR-theorem starts with two crucial assumptions:

\begin{itemize}
\item[-]
\textbf{Assumption 1 (Reality Criterion)}: Every physical system has an underlying ontic state that completely describes its physical properties, which is objective and independent from observers.\footnote{One may think that in order to assign an ontic state to a physical system, one needs to presuppose that the system is sufficiently isolated from the environment. For example, if one entangles two electrons in the singlet state, neither electron has an ontic state, because we can't assign a (pure) wave-function to either electron. I think this example confuses the ontic state $\lambda$ with the wave-function $\psi$, which is supposed to describe or represent properties of the ontic state. The ontic state is independent of a physical theory, and it captures all the physical properties of the system. Therefore, both electrons have their own ontic state. The ontological model framework investigates whether a pure-state wave-function gives a complete or partial physical description of the ontic state of the system. According to quantum mechanics, neither electron in the singlet state can be assigned a pure-state wave-function, but this does not mean that there is no complete physical description of them. One may criticize the ontological models framework, however, that it ignores density matrices as representations of the ontic state \citep{Carcassi:2022uu}. But this criticism is rather about the representation of $\lambda$ than of $\lambda$ itself. It would be a worthwhile project to investigate the implications for density matrix realism \citep{Chen:2021tq}, if one generalizes the ontological models framework and the PBR theorem. (I thank
Charles Sebens and Eddy Chen 
for helpful discussions on these issues.)}
\item[-]
\textbf{Assumption 2 (Preparation Independence)}: Two preparation devices run independently from each other.
\end{itemize}
The first assumption is essential to the ontological models framework, whereas the second is added for methodological reasons.
The reality criterion basically says that we consider a theory that fits into the ontological models framework. Preparation independence adds to the ontological models framework that the statistical distributions of ontic states that are generated in one preparation device do not depend on  the statistical distributions of ontic states that are generated in another preparation device.\footnote{One may reason that the preparation devices need to be space-like separated to run independently. That is just one way to justify that they have to run independently. Even if the preparation devices are not space-like separated assumption $2$ may still hold if there is no causal relation between the devices.}

The general argument of the PBR-theorem is the following: if the wave-function is $\psi$-epistemic and the two assumptions hold, then we get a contradiction with the (well-confirmed) predictions of quantum mechanics. Since the two assumptions are reasonable, the wave-function cannot be $\psi$-epistemic and must be therefore $\psi$-ontic. It is indeed possible to question the validity of the two assumptions and explore how such a quantum theory would look like.\footnote{Theories that deny the reality assumption are QBism \citep{Fuchs:2017aa,Fuchs:2014ab}, radical epistemicism \citep{Ben-Menahem:2017aa,Ben-Menahem:2018ur,Ben-Menahem:2020aa}, relational quantum mechanics \citep{Rovelli:1996wy, Oldofredi:2020wv, Oldofredi:2021ta, Di-Biagio:2021vt}, and pragmatist interpretations of quantum mechanics \citep{Healey:2017um}. Theories that deny preparation independence are Spekken's toy model \citep{Spekkens:2007vg} and different proposals for superdeterministic theories \citep{Palmer:1995aa,Hooft:2016aa,Hossenfelder:2019aa,Ciepielewski:2021vn}. An excellent critical review of this approach is \citet{Chen:2021vv}. } I just want to emphasize here that it is impossible to violate the Reality Criterion and retain a $\psi$-epistemic wave-function, as is often claimed. For the definition of $\psi$-epistemic hinges on the ontological models framework and thus on the existence of $\lambda$.

\begin{figure}[ht]
  \centering
 \includegraphics[width=10cm]{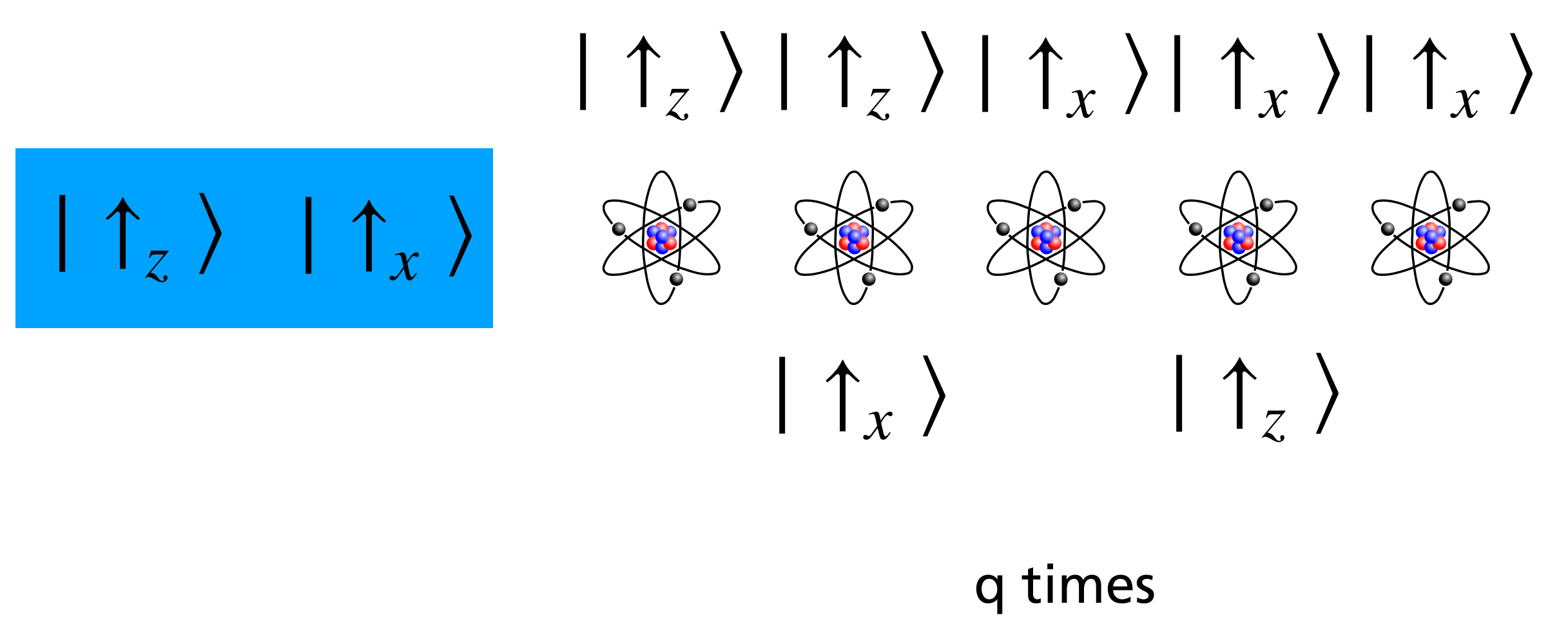}
\caption{The preparation of systems in the PBR-theorem. If the wave-function is $\psi$-epistemic, some of the prepared system can are associated with two different wave-functions. It is assumed that this happens $q$ times.}
\label{fig:state-preparation}
\end{figure}

Let us now discuss the experimental set-up of the PBR-theorem. Two preparation devices are used that generate particles either in a $z$-spin up or an $x$-spin up state (see Fig.\ \ref{fig:state-preparation}).  
Alice and Bob can choose in which state the system should be prepared. Since it is assumed that the wave-function is $\psi$-epistemic some of the wave-functions in Alice's ensemble can be correctly associated with another wave-function. For simplicity, we assume that if such a system is in an $z$-spin up state the other correct wave-function would be $x$-spin up and vice versa. The same is the case for Bob's systems. To have a concrete number, we say that these ontic states with double wave-functions appear $q$ times in such an ensemble. 

 \begin{figure}[ht]
  \centering
 \includegraphics[width=15cm]{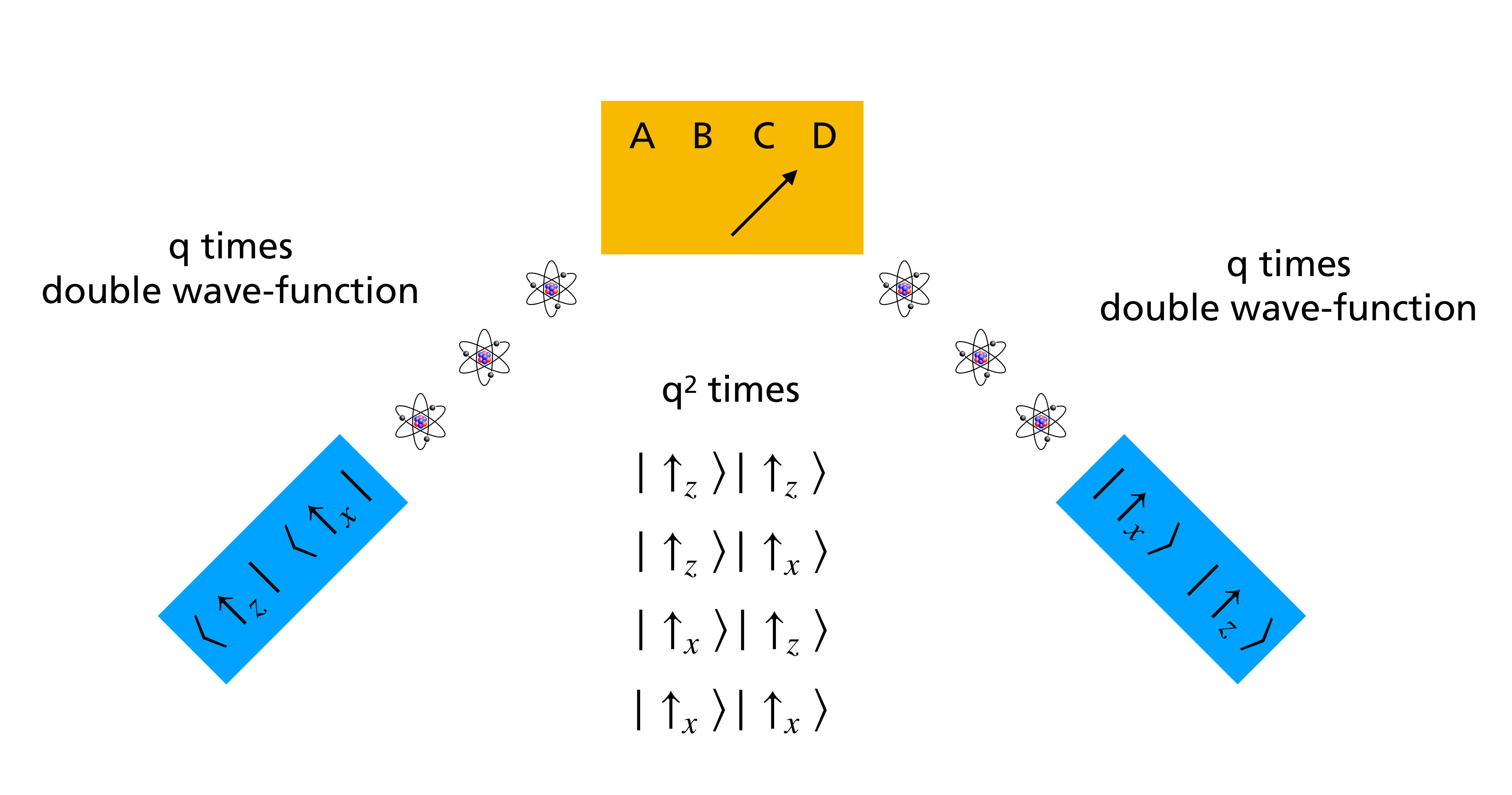}
\caption{The proof of the PBR-theorem. Alice and Bob prepare an ensemble of quantum system that are either in a $z$-spin up or and $x$-spin up state. They send their system to a measurement device (orange box) that makes a measurement on pairs of systems, one system in such a pair is from Alice's and one from Bob's preparation device.}
\label{fig:pbr-proof}
\end{figure}

Now Alice and Bob shoot their quantum systems into a measuring device that is set up to measure pairs of particles, where one system of the pair comes from Alice and the other from Bob. Certain pairs are of particular interest, because these are the ones that ultimately yield a contradiction with the predictions of quantum mechanics. A fraction of $q$ systems in Alice's and on Bob's side are associated with two wave-function: $\vert \uparrow_z \rangle$ and $\vert \uparrow_x \rangle$. Due to preparation independence, the ontic states  with such double wave-functions on Alice's side are independent from (or uncorrelated with) the ontic states with double wave-functions on Bob's side. Therefore, $q^2$ times a pair of particles is associated with four wave-functions (each in a product state): $\vert \uparrow_z \rangle \vert \uparrow_z \rangle$,  $\vert \uparrow_z \rangle \vert \uparrow_x \rangle$, $\vert \uparrow_x \rangle \vert \uparrow_z \rangle$, and $\vert \uparrow_x \rangle \vert \uparrow_x \rangle$.

The subsequent measurement is set up in such a way to yield one of four possible measurement outcomes: $A$, $B$, $C$, or $D$. What are the probabilities for either outcome if a pair in a product state is measured? The device is prepared that a  $\vert \uparrow_z \rangle \vert \uparrow_z \rangle$ state will not get measured as $A$, that is, the probability for yielding $A$ given this product state is zero: $P_{z,z}(A)=0$. Similarly, for the other product states: $P_{z,x}(B)=0$, $P_{x,z}(C)=0$, $P_{x,x}(D)=0$.\footnote{For our purposes it is not important to explain the details of this measurement procedure. The product states get projected onto a certain entanglement basis that gives rise to these probabilities.} It is now easy to see why these ontic states that are associated with all these product states are problematic. If we shoot these states into the measurement device, the theory tells us that the measurement will not give us either of the four possible results. On the other hand, the measurement apparatus by construction will measure something and will yield one of the four possible outcomes.\footnote{The precise argument why the measurement apparatus has to show a result is a bit technical. Roughly speaking it goes like this. We deal here with a four-dimensional Hilbert space. Each outcome of the measurement device is associated with a basis vector in this Hilbert space. These basis vectors span the \emph{complete} Hilbert space. So upon measurement any vector in this Hilbert space gets projected on one of these basis vectors and yields the corresponding value with a certain probability. In particular, the product states $\vert \uparrow_z \rangle \vert \uparrow_z \rangle$,  $\vert \uparrow_z \rangle \vert \uparrow_x \rangle$, $\vert \uparrow_x \rangle \vert \uparrow_z \rangle$, and $\vert \uparrow_x \rangle \vert \uparrow_x \rangle$ are part of this Hilbert space. So it is impossible that the measurement device will yield nothing if fed with one of the problematic ontic states.} Overall, it happens $q^2$ times that the measurement device will show a result that is prohibited by quantum mechanics if the wave-function is $\psi$-epistemic.

One can now react in three different ways to the result of the PBR-theorem:
\begin{enumerate}
\item
Our best choice of wave-functions is to interpret them as $\psi$-ontic. 
\item
We may deny the preparation independence assumption. These theories would fit into the ontological models framework, they would also have a $\psi$-epistemic wave-function, but they would not lead to the contradiction indicated in the PBR-theorem.\footnote{It is, in principle, a further option to have a $\psi$-ontic wave-function and deny preparation independence. \citet{Ciepielewski:2021vn} present such a model, although they do not defend it.}
\item
We may deny the reality criterion. Then the entire set up of the ontological models framework would be undermined; in particular, the PBR-theorem would say nothing about these kinds of quantum theories. 
\end{enumerate}

I do not want to argue for or against either of these strategies in this paper. Instead, this overview will give us a scheme for evaluating the status of the wave-function in the statistical interpretation of quantum mechanics. 

\section{The Statistical Interpretation of Quantum Mechanics}
\label{sec:stat-int-ontic}


For \citet[][]{Harrigan:2010aa}, the wave-function in the statistical interpretation is $\psi$-epistemic and $\psi$-incomplete, while \citet[][]{Oldofredi:2020wv} claim it to be  $\psi$-ontic and $\psi$-complete. I will argue that Einstein and Ballentine regard the wave-function in the statistical interpretation as $\psi$-epistemic and $\psi$-incomplete. I will also discuss whether the wave-function in the statistical interpretation can be $\psi$-statistical. I will conclude, however, that the wave-function in the statistical interpretation needs to be $\psi$-ontic and $\psi$-incomplete.

Let us first start to introduce the statistical interpretation of quantum mechanics. One of the clearest descriptions of it can be found in Einstein's 1949 \emph{Reply to Critics} in Paul Schilpp's volume \emph{Albert Einstein: Philosopher--Scientist}:

\begin{quotation}
Within the framework of statistical quantum theory there is no such thing as a complete description of the individual system. More cautiously it might be put as follows: The attempt to conceive the quantum-theoretical description as the complete description of the individual systems leads to unnatural theoretical interpretations, which become immediately unnecessary if one accepts the interpretation that the description refers to ensembles of systems and not to individual systems. In that case the whole ‘egg-walking’ performed in order to avoid the ‘physically real’ becomes superfluous. There exists, however, a simple psychological reason for the fact that this most nearly obvious interpretation is being shunned. For if the statistical quantum theory does not pretend to describe the individual system (and its development in time) completely, it appears unavoidable to look elsewhere for a complete description of the individual system; in doing so it would be clear from the very beginning that the elements of such a description are not contained within the conceptual scheme of the statistical quantum theory. With this one would admit that, in principle, this scheme could not serve as the basis of theoretical physics. Assuming the success of efforts to accomplish a complete physical description, the statistical quantum theory would, within the framework of future physics, take an approximately analogous position to the statistical mechanics within the framework of classical mechanics. I am rather firmly convinced that the development of theoretical physics will be of this type; but the path will be lengthy and difficult. \citep[][pp.\ 671--672]{Schilpp:1970aa}
\end{quotation}

I want to take home three important points from this passage:
\begin{enumerate}
\item
The wave-function in the statistical interpretation describes ensembles instead of individual quantum systems.
\item
Therefore individual systems are not completely described by the wave-function. 
\item
To describe an individual system completely, one would need to go beyond the statistical interpretation to search for such a completion.
\end{enumerate}
The statistical interpretation, as Einstein describes it, is a peculiar form of interpretation. Normally, one would seek an interpretation of quantum mechanics that is in some sense complete. Even if one were to support an operational or pragmatist interpretation of quantum mechanics \citep{Healey:2017um}, one would argue that such an interpretation does not need a completion---for example, because one is skeptical that we can discover the true nature of unobservable objects. The statistical interpretation, on the other hand, rather says that, \emph{for all practical purposes}, one can think of the wave-function describing ensembles, but this is not the complete story of reality. In his last words on the statistical interpretation, Einstein emphasized in a letter to A.\ Lamouche just a month before he died on March 20, 1955, that the statistical interpretation is incomplete:
\begin{quote}
The $\psi$-function is not to be considered as a complete description of an individual state of affairs, rather only as a representation of what we can know about a particular state of affairs from an empirical point of view. Then the $\psi$-function is a representation of an ``ensemble'', not the complete characterization of individual states of affairs. One has thereby renounced the latter in principle.\footnote{I thank 
Maaneli Derakhshani 
for this reference.}  
 $\,$ \citep[quoted in][p.\ 9]{Fine:1993um}
\end{quote}

We can find the same attitude in \citet{Ballentine:1970aa,Ballentine:1972vt,Ballentine:2015aa}, who is probably the most famous modern advocate of the statistical interpretation:

\begin{quote}
The Statistical Interpretation, according to which a pure state provides a description of certain statistical properties of an ensemble of similarly prepared systems, but need not provide a complete description of an individual system. \citep[][p.\ 360]{Ballentine:1970aa}
\end{quote}

\begin{quote}
We see that a quantum state is a mathematical representation of the result of a certain state preparation procedure. Physical systems that have been subjected to the same state preparation will be similar in  some of their properties, but not in all of them […]. \citep[][p.\ 361]{Ballentine:1970aa}
\end{quote}

\begin{quote}
The Statistical Interpretation, which regards quantum states as being descriptive of ensembles of similarly prepared systems, is completely open with respect to hidden variables. It does not demand them, but it makes the search for them entirely reasonable [(]this was the attitude of Einstein (1949)[)]. \citep[][p.\ 374]{Ballentine:1970aa}
\end{quote}

\begin{quote}
The Statistical Interpretation does not prejudice the possibility of introducing hidden variables which would determine (in principle) the outcome of each individual measurement (Sec. 6). \citep[][p.\ 379]{Ballentine:1970aa}
\end{quote}

For both Einstein and Ballentine, the wave-function does indeed give an incomplete description of physical systems and would need a completion to provide such a description, as Ballentine's statements ``The Statistical Interpretation does not prejudice the possibility of introducing hidden variables'' and that the statistical interpretation ``is completely open with respect to hidden variables'' confirm. On the one hand, it seems that Ballentine regards the statistical interpretation as a viable interpretation for all practical purposes, but, on the other hand, this interpretation is incomplete and can be completed by hidden variables. The statistical interpretation is a useful interpretation of quantum mechanics, especially for physicists, and it gives a clear statistical interpretation of the predictions of quantum mechanics. This interpretation does therefore provide a richer picture of the world than pure operationalism and a less obscure metaphysics than the Copenhagen interpretation, even if the wave-function is $\psi$-incomplete.

\subsection{$\psi$-epistemic and incomplete?}

Is the wave-function in this interpretation $\psi$-ontic or $\psi$-epistemic? Einstein does not  explicitly answer this question in the above quotation. But since the wave-function describes ensembles instead of individual quantum systems, it seems plausible that an individual system can be associated with two different wave-functions. Einstein mentions such a case in his 1935 correspondence with Schrödinger (\citealp[see][section 2]{Howard:1985wa} and \citealp[][section 4.3]{Harrigan:2010aa}):

\begin{quotation}
Now what is essential is exclusively that $\psi_{B} $ and $\psi_{\underset{\bar{}}{B}}$ are in general different from one another. I assert that this difference is incompatible with the hypothesis that the $\psi$
description is correlated one-to-one with the physical reality (the real state). After
the collision, the real state of $(AB)$ consists precisely of the real state of $A$ and the
real state of $B$, which two states have nothing to do with one another. \emph{The real state
of $B$ thus cannot depend upon the kind of measurement I carry out on $A$} [\dots] But then for the same state of $B$ there are two (in general
arbitrarily many) equally justified $\psi_B$, which contradicts the hypothesis of a one-to-one
or complete description of the real states. \citep[Einstein to Schrödinger in 1935, quoted in][p.\ 180]{Howard:1985wa}
\end{quotation}
Einstein concocted several arguments throughout his career to prove that quantum mechanics is incomplete. The only option Einstein saw in making sense of quantum mechanics was to interpret it as an incomplete statistical theory. So when Einstein concludes that quantum mechanics is incomplete, we can also understand this to say that the statistical interpretation is incomplete. His arguments also show what this incompleteness amounts to. In the above quote, Einstein talks about an entangled two-particle system $AB$, presumably in this state: $\psi=\frac{1}{\sqrt{2}}\left(\psi_A\psi_B+\psi_{\underset{\bar{}}{A}}\psi_{\underset{\bar{}}{B}}\right)$. We can think of two electrons in the singlet state, for example. Einstein presupposed that each particle is prepared in an ontic state $\lambda_{A}$ and $\lambda_{B}$ respectively. Whether or not we measure particle $A$, this measurement cannot have any physical influence on the other particle, since Einstein strongly believed in locality. Therefore, the ontic state of particle $B$ remains the same before and after measurement, and the wave-function representing this state cannot change either. Thus, the ontic state $\lambda_{B}$ is correctly described by both $\psi_{B} $ and $\psi_{\underset{\bar{}}{B}}$. Thus quantum mechanics, as well as the statistical interpretation, is $\psi$-epistemic and $\psi$-incomplete.\footnote{\citet{Fano:2019wp} contest that this is Einstein's line of argument. According to them, Einstein argued only that if locality holds standard quantum mechanics must be $\psi$-epistemic. It's not so clear whether Einstein had a coherent view of what the statistical interpretation actually is. Sometimes Einstein can be read to advocate a $\psi$-ontic interpretation but also a $\psi$-epistemic interpretation in other contexts.} Einstein deduces the incompleteness from the issue that several wave-functions can be associated with the same ontic state; a complete theory, on the other hand, would uniquely associate a wave-function to an ontic state. \citeauthor{Harrigan:2010aa} also interpret this passage of Einstein's as arguing for a $\psi$-epistemic interpretation of the wave-function. They conclude: 
\begin{quotation}
By characterizing his 1935 argument as one that merely established the \emph{incompleteness}
of quantum theory on the assumption of locality, Einstein did it a great disservice.
For in isolation, a call for the \emph{completion} of quantum theory would naturally
have led many to pursue hidden variable theories that interpreted the fundamental
mathematical object of the theory, the wave function, in the same manner in which
the fundamental object of other physical theories were customarily treated—as ontic.
But such a strategy was known by Einstein to be unable to preserve locality. Thus
it is likely that the force of Einstein’s 1935 argument from locality to the epistemic
interpretation of $\psi$ was not felt simply because the argument was not sufficiently well
articulated. \citep[][p.\ 152]{Harrigan:2010aa}
\end{quotation}

Ballentine's answer as to whether the statistical interpretation is $\psi$-epistemic or $\psi$-ontic is encrypted in his second quote: ``Physical systems that have been subjected to the same state preparation will be similar in  some of their properties, but not in all of them.'' This sounds like he regards the wave-function as $\psi$-statistical. An individual system has a certain wave-function solely in virtue of being part of an ensemble that has been prepared in the same quantum state, \emph{but some of these systems within this ensemble may differ with respect to their physical properties.} If these systems can be described by another wave-function, they would have a $\psi$-statistical wave-function, and then the wave-function would be $\psi$-epistemic. 

If Ballentine would advocate for such a statistical interpretation, it would not be able to make correct empirical predictions according to the PBR theorem. The only way out would be a $\psi$-epistemic completion of the statistical interpretation that violates the assumption of preparation independence, which would then lead to some form of super-determinism \citep[see][for such a model]{Spekkens:2007vg,Leifer:2014wv}.\footnote{Ballentine could also deny the reality criterion (I thank an anonymous reviewer for raising this option). Then the statistical interpretation would not fit into the ontological models framework; it would be neither $\psi$-ontic nor $\psi$-epistemic; and the PBR theorem would not be applicable. The biggest problem I see here is that it becomes unclear what the statistical interpretation is about and how the statistical pattern is generated in the first place. If there is no ontological underpinning in terms of some objective properties, it is not obvious what a quantum system is and how this (non-existing?) quantum system can interact with a measurement device. \citet{Oldofredi:2020wv} will take a similar but less radical approach by revising instead of abolishing the reality criterion (see section \ref{subsec:ontic-complete}).}

Let me briefly explain why one needs to violate preparation independence to rescue $\psi$-epistemic wave-functions.\footnote{In their analysis of the statistical interpretation, \citet{Fano:2019wp} interpret the PBR-theorem to make such an interpretation impossible. Violation of preparation is in fact a way, although problematic, to retain a $\psi$-epstemic wave-function.} Remember the PBR setting in Fig.\ \ref{fig:pbr-proof} from section \ref{sec:pbr-theorem}. Two preparation devices (blue boxes) send a beam of particles to a measurement device (orange box). The particles are either prepared in a spin $x$-up or spin $z$-up state. Since it is assumed that the wave-function is $\psi$-epistemic, a fraction $q$ of the particles in each beam are described by both the spin $x$-up and spin $z$-up wave-functions (double wave-functions). Due to preparation independence, the ontic states of the left beam are independently distributed from the ontic states in the right beam; therefore, $q^2$ of times, a pair of particles (one from the left and one from the right beam) are associated with \emph{four} wave-functions. These simultaneous four wave-functions lead to a contradiction with the predictions of quantum mechanics. A violation of preparation independence would not yield any two-particle system that is correctly described by these four wave-functions. That is, whenever a particle in the left beam has a double wave-function the corresponding particle in the right beam will only have one correct wave-function, and vice versa. Thus, the distribution of double wave-functions in each beam are correlated such that a particle with a double wave-function in one beam will not be paired with a particle in the other beam having also a double wave-function.

There are a couple of problems with violating preparation independence. First, how could the ontic states of the two beams be correlated in the first place? This becomes particularly problematic when we isolate each preparation device from each other---for example, by space-like separation and special isolation materials round them. There are two ways to do that, which are both implausible. Either these correlations happen by pure chance. If that were the case, there needs to be some case (even only in principle) where chance is not in our favor and would match two systems with a double wave-function. This would, however, undermine the empirical predictions of the theory. Another way to explain these correlations would be to postulate special fine-tuned initial conditions in the past before the preparation (and the measurement) were conducted. These fine-tuned initial conditions are ultimately traced back to  fine-tuned initial conditions of the universe\dots and these fine-tuned initial conditions demand further explanation. Especially,  when experimentalists do not do anything particularly special with the preparation devices, such special initial conditions have a mysterious character. 

Second, the correlation of the ontic states after preparation is sensitive to the future measurement. If we change the measurement device and conduct a different kind of measurement, this would necessitate that the particles become differently correlated if this measurement is similarly prepared as in the PBR-theorem---otherwise, we would run to another contradiction. One may explain the influence of the measurement device again by special initial conditions, but this would make the set-up even more fine-tuned. And if the future measurement is not specially designed as in the PBR-theorem, does this mean that the particles would be still correlated or are they uncorrelated?

Third, connected to the previous point, a violation of preparation independence depends on operational procedures, like measurement and preparation. It is unclear where to draw the line between measurement and non-measurement procedures and between preparation and non-preparation procedures. It is therefore ill-defined in which situations one has correlations and in which one does not. Peter \citet{Lewis:2006ab} calls this a measurement-problem-like problem. 

If the statistical interpretation is $\psi$-epistemic and incomplete as claimed by \citet{Harrigan:2010aa}, it has to violate preparation independence and rely on some super-deterministic mechanism, otherwise it would not be an empirically adequate theory. To avoid a violation of preparation independence, you may seek an interpretation of the wave-function as $\psi$-ontic. \citet[][]{Oldofredi:2020wv} go along this route but for different reasons. I present and evaluate their arguments in the next subsection. 

\subsection{$\psi$-ontic and complete?}
\label{subsec:ontic-complete}
The biggest problem \citet[][]{Oldofredi:2020wv} identify in \citeauthor{Harrigan:2010aa}'s classification of the statistical interpretation is that this interpretation demands a different kind of ontic state:

\begin{quote}
In the second place, another crucial point to highlight is that the ontic space of the statistical interpretation is not one of individuals, but of ensembles. This allows for an alternative reading of the ontic state: it provides a complete description of the properties of an ensemble, not of individuals. And there is nothing else to know about ensembles that is not provided by the quantum state. The upshot of the present discussion is that the sort of $\lambda$ that the statistical interpretation poses is completely different in nature with respect to that employed by Harrigan and Spekkens. \citep[][p.\ 1330]{Oldofredi:2020wv}
\end{quote}
According to \citet[][]{Oldofredi:2020wv}, the ontic state that underlies the statistical interpretation is not the one that is presupposed in the ontological models framework (which refers to individual systems), but one that only refers to an ensemble of systems. With this kind of ensemble-$\lambda$, one may then interpret the wave-function in the statistical interpretation to be $\psi$-ontic and also $\psi$-complete. Since, in the usual definition, a $\psi$-ontic wave-function requires an ontic state for individual systems, \citet[][]{Oldofredi:2020wv} would need an ``ensemble ontological model'', in which the wave-function uniquely refers to the ensemble-$\lambda$.

\citet[][]{Oldofredi:2020wv} do not argue that a $\psi$-epistemic wave-function is troublesome because of  the PBR-theorem, that is, there are physical reasons to construe the wave-function differently; rather, they mention historical reasons for their take on the wave-function and the corresponding ontic state. They argue that Ballentine thought the wave-function in the statistical interpretation to provide a complete description of physical systems. Since any $\psi$-complete wave-function is necessarily $\psi$-ontic \citep[see Lemma 6 in][p.\ 133]{Harrigan:2010aa}, the wave-function in the statistical interpretation is $\psi$-ontic.

Their historic argument is quite confusing. First, they present three sources in which Einstein explicitly argues that the wave-function has to be \emph{incomplete} (to not violate locality).\footnote{They are (i) Einstein's remarks at the 1927 Solvay conference \citep[transcript in][pp.\ 440--442]{Bacciagaluppi:2009aa}, (ii) Einstein's 1936 essay \emph{Physics and Reality} \citep{Einstein:1936tz}, and (iii) Einstein's reply to critics in his intellectual autobiography \citep{Schilpp:1970aa}.} Then they say that ``Hence, it is fair to establish a strong theoretical continuity between Ballentine’s
presentation of the ensemble view and Einstein’s interpretation of quantum
mechanics.'' (p.\ 1330) That is, Ballentine agrees with Einstein on what the statistical interpretation tells us about quantum systems \citep[see also][]{Ballentine:1972vt}.\footnote{\citet{Fine:1993um} also discusses how Einstein interpreted quantum mechanics.} In particular, Ballentine is supposed to agree with Einstein that this theory is incomplete. As I discussed above, Ballentine sometimes appears to be indecisive as to whether his interpretation is complete or incomplete, but he certainly considers that his theory can be completed by another quantum theory, but whether it needs such a completion is left open. \citeauthor{Oldofredi:2020wv} seem to suggest that Ballentine breaks with Einstein and advocates an interpretation that is complete necessitating a different ontic state $\lambda$ that only refers to entire ensemble not to individual systems. 

In my reading of Ballentine, I rather think that he considers his theory ``complete for all practical purposes'' to make successful predictions and to apply quantum mechanics without dealing with a mysterious metaphysics à la Copenhagen, but to get a truly complete theory that tells us what quantum systems really are, Ballentine seems to embrace such a completion of the statistical interpretation. 

Hence, it is \citeauthor{Oldofredi:2020wv}'s proposal of an ensemble-$\lambda$, which renders the statistical interpretation a complete theory,  and therefore it is them who break with both Einstein's and Ballentine's view of how the statistical interpretation refers to the physical world.\footnote{I also think that \citeauthor{Oldofredi:2020wv} cite the wrong reason for why \citeauthor{Harrigan:2010aa} consider the statistical interpretation to be $\psi$-epistemic: 
\begin{quotation}
%
How do Harrigan and Spekkens reinforce
their conclusion that the ensemble view is a $\psi$-epistemic model? They answer it by
saying that the notion of `ensemble'' in Einstein’s jargon is nothing but a way to talk
about probabilities reflecting an observer’s knowledge[.] \citep[][p.\ 1328]{Oldofredi:2020wv}
\end{quotation}
\citeauthor{Harrigan:2010aa} do not make the mistake of identifying $\psi$-epistemic wave-functions with epistemic interpretations of the wave-function. Instead, they conclude that the wave-function in the statistical interpretation to be $\psi$-epistemic on two grounds. First, it is incomplete (according to Einstein and Ballentine). Second, Einstein gives an argument of associating two different wave-function to the same quantum system in his 1935 correspondence to Schrödinger.  
}
I see the following problems with their suggestion. First, as I said, Einstein and Ballentine explicitly mention that their interpretation is incomplete. For all practical purposes, one can use the statistical interpretation to have a sufficiently clear picture of what quantum physics tells us about the world, but for a complete picture one would need to supplement this interpretation. Second, the statistical behavior of an ensemble is generated by its individual constituents. A natural question is how these constituents do that. How is the ensemble-$\lambda$ related to the individual $\lambda$s? Third, and connected to the previous point, even if the wave-function provides the complete description of an ensemble, it is unclear how the wave-function could not be $\psi$-statistical. Such a statistical interpretation of the wave-function would not provide the means to exclude that some particles can be swapped between two ensembles. One may respond that it is impossible to talk about the individual ontic states since all there is in this statistical interpretation is the ensemble-$\lambda$. This response is problematic, because the individual systems do exist, and it is unclear why we are not supposed to completely describe them or why the properties of the individual systems do not contribute to the complete description of the ensemble. Fourth, ensembles are usually defined to be series or copies of infinitely many systems. In the real world, we only deal with finite series. It is not clear how an ontic state that is about the properties of infinitely many systems is related to the properties of finitely many systems. 

\subsection{$\psi$-ontic and incomplete!}
Because of the just mentioned conceptual problems with an $\psi$-ontic and complete wave-function for the statistical interpretation and the challenges of the PBR-theorem faced by a $\psi$-epistemic and incomplete wave-function, I conclude that a $\psi$-ontic and incomplete wave-function is the best option. 

One may argue that a $\psi$-ontic and incomplete wave-function would restrict the relation between the ontic state $\lambda$ and $\psi$ more strictly than the formulation of the statistical interpretation justifies, because although the wave-function in the statistical interpretation only describes ensembles, we would need to believe in (a not yet specified) completion that would describe \emph{individual systems}. This description of individual systems would make it impossible that an electron can be swapped from an $x$-up beam into a $z$-up beam. Even if the statistical interpretation does not seem to a priori prohibit such a swap of particles, this constraint follows from the wave-function being $\psi$-ontic. That is, a particle is in a certain quantum state $\psi$ not because it happens to be prepared to be part of an ensemble, but because properties of its ontic state $\lambda$ make it to be in the quantum state $\psi$. 

Although it is correct that the statistical interpretations allows for different ways the wave-function could be, I presented several arguments that uncover the problems of a $\psi$-epistemic or a complete wave-function. These arguments are similar in kind to arguments that have been made about the contextuality \citep{Kochen:1967aa} and non-locality of quantum mechanics \citep{Bell:1964wu}. The Kochen-Specker theorem concludes that a $\psi$-incomplete wave-function can only be supplemented by contextual variables (that is, variables that lead to different empirical results depending on how they are measured); Bell's theorem says that as long as a certain kind of statistical independence is fulfilled the theory cannot explain certain correlations by a local mechanism. So even if a certain interpretation of quantum mechanics is incomplete, one can discover certain hidden structure of the theory or certain features of a possible completion. The same is the case for my defense of a $\psi$-ontic and $\psi$-incomplete wave-function for the statistical interpretation. 


This reading of the statistical interpretation would indeed break with Einstein's view of assigning several wave-functions to the same system but seems to back up my understanding of Ballentine of using the statistical interpretation for all practical purposes, while being open to a possible completion. Such an interpretation would also make the statistical interpretation fit into the ontological models framework without a revisionary interpretation of the ontic state $\lambda$.


\section{Conclusion}
Is the statistical interpretation of quantum mechanics $\psi$-epistemic? It could be.  The core tenets of the interpretation do not settle the question as to whether the wave function is $\psi$-epistemic or $\psi$-ontic.  That is left open, to be decided by deeper physics. If the statistical interpretation allows for the swap of some particles prepared with different wave-functions, the wave-function would be $\psi$-epistemic, but then it would be ruled out by the PBR theorem, unless one is willing to violate preparation independence and support a quantum theory that invokes some form of super-determinism. The wave-function would be $\psi$-onic, if the statistical interpretation is completed by a quantum theory that describes the state of individual systems. 

The way the statistical interpretation is formulated by \citeauthor{Ballentine:1970aa} would paradoxically make both the $\psi$-epistemic and the $\psi$-ontic path viable. The underlying issue is that this interpretation in its very definition is incomplete.

\section*{Acknowledgements}
I wish to thank Gabriele Carcassi, Eddy Chen, Maaneli Derakhshani, Joshua Eisenthal, Christopher Hitchcock, Dustin Lazarovici, Matthew Leifer, Cristian López, Travis Norsen, Andrea Oldofredi, and Charles Sebens for very helpful comments and discussions on a previous draft of this paper. Special thanks go to Tim Maudlin for many hours of invaluable discussions on this topic. Lastly, I want to thank an anonymous reviewer for helpful comments.

\bibliographystyle{abbrvnat}
\bibliography{references}
\end{document}